\documentclass[twocolumn,showpacs,preprintnumbers,amsmath,amssymb,superscriptaddress,floatfix]{revtex4-2}
\usepackage{epsf}
\usepackage{float}
\usepackage{graphicx}
\usepackage{sidecap}
\usepackage[utf8]{inputenc}
 \usepackage{soul}
 \usepackage{array}
 \usepackage{amsmath}
 \usepackage{amssymb}
 \usepackage{dcolumn}
 \usepackage{epstopdf}
 \usepackage{bm}

 \usepackage{amsmath}
 
\usepackage{hyperref}

\sethlcolor{yellow}
\hypersetup{
    colorlinks=true,
    citecolor=red,
    linkcolor=red,
    filecolor=blue,   
    urlcolor=blue,
}

\begin{document}

\title{Staged emergence of anomalous Hall transport in a correlated uranium Weyl semimetal}
\author{Sabin~Regmi}\thanks{\href{mailto:sabin.regmi@inl.gov}{sabin.regmi@inl.gov}} \affiliation{Center for Quantum Actinide Science and Technology, Idaho National Laboratory, Idaho Falls, Idaho 83415, USA}
\author{Shuxiang~Zhou} \affiliation{Center for Quantum Actinide Science and Technology, Idaho National Laboratory, Idaho Falls, Idaho 83415, USA}
\author{Chandan K. Singh} \affiliation{Department of Physics and Astronomy, Uppsala University, P.O. Box 516, SE-75120 Uppsala, Sweden}
\author{Alexei~Fedorov} \affiliation{Advanced Light Source, Lawrence Berkeley National Laboratory, Berkeley, California 94720, USA}
\author{Jonathan~Denlinger} \affiliation{Advanced Light Source, Lawrence Berkeley National Laboratory, Berkeley, California 94720, USA}
\author{Zeyu~Ma} \affiliation{School of Engineering \& Applied Sciences, Harvard University, Cambridge, MA 02138, USA}
\author{Yidi~Wang} \affiliation{Department of Physics, Harvard University, Cambridge, MA 02138, USA}
\author{Jennifer~E.~Hoffman} \affiliation{School of Engineering \& Applied Sciences, Harvard University, Cambridge, MA 02138, USA} \affiliation{Department of Physics, Harvard University, Cambridge, MA 02138, USA}
\author{Peter M. Oppeneer} \affiliation{Department of Physics and Astronomy, Uppsala University, P.O. Box 516, SE-75120 Uppsala, Sweden}
\author{Dariusz~Kaczorowski} \affiliation{Institute of Low Temperature and Structure Research, Polish Academy of Sciences, Okólna 2, 50-422 Wrocław, Poland}
\author{Tomasz~Durakiewicz}\affiliation{Center for Quantum Actinide Science and Technology, Idaho National Laboratory, Idaho Falls, Idaho 83415, USA}
\author{Krzysztof~Gofryk}\thanks{\href{mailto:gofryk@inl.gov}{gofryk@inl.gov}} \affiliation{Center for Quantum Actinide Science and Technology, Idaho National Laboratory, Idaho Falls, Idaho 83415, USA} \affiliation{Glenn T. Seaborg Institute, Idaho National Laboratory, Idaho Falls, Idaho 83415, USA}

\begin{abstract}

Understanding how electronic correlations modify topological electronic states remains a central challenge in quantum materials. Uranium compounds provide a particularly attractive platform because their spatially extended $5f$ orbitals combine strong spin-orbit coupling, substantial hybridization, and intermediate electronic correlations on comparable energy scales. Here we investigate the uranium ferromagnet UPS using magnetotransport, angle-resolved photoemission spectroscopy (ARPES), thermodynamic measurements, and first-principles calculations. Resonant ARPES reveals the coexistence of narrow U-$5f$ spectral weight at the Fermi level with broad incoherent electronic states, demonstrating the dual itinerant and correlated nature of the uranium $5f$ electrons. The Hall response develops progressively across the ferromagnetic transition and evolves over a substantially broader temperature range than the magnetic order parameter. The anomalous Hall conductivity reaches approximately $4.5\times10^{2}\,\Omega^{-1}\,\mathrm{cm}^{-1}$, yet its temperature dependence does not simply follow the ordered magnetic moment. Instead, temperature-dependent ARPES results reveal a redistribution of low-energy $5f$ spectral weight below approximately 90~K, showing that the correlated electronic structure continues to evolve well inside the ferromagnetic state. First-principles calculations identify a symmetry-protected Weyl crossing with pronounced Berry curvature and yield an intrinsic anomalous Hall conductivity of approximately $9.6\times10^{2}\,\Omega^{-1}\mathrm{cm}^{-1}$ at the calculated Fermi level, of the same order of magnitude as experiment. Together, these results demonstrate that magnetic order, anomalous Hall transport, and correlated $5f$ electronic states develop over distinct but overlapping temperature ranges, revealing a staged electronic reconstruction in a correlated uranium Weyl semimetal.

\end{abstract}

\maketitle

\textit{Introduction} - The interplay between strong electronic correlations and band topology has emerged as one of the central themes in condensed matter physics. While Berry curvature and topological transport phenomena are now well understood in weakly correlated topological semimetals, considerably less is known about their evolution in systems where the low-energy electronic structure itself develops through many-body interactions. In correlated materials, quasiparticle formation, magnetic order, and anomalous Hall transport need not emerge simultaneously. Instead, electronic correlations may progressively reconstruct the low-energy electronic structure, modify the Berry-curvature landscape, and produce transport responses that evolve over multiple characteristic energy scales rather than at a single thermodynamic phase transition \cite{Nagaosa2010,Burkov2018,Armitage2018}. This raises a fundamental question: Does anomalous Hall transport emerge simultaneously with magnetic order, or does it instead reflect the progressive reconstruction of correlated electronic states?

\begin{figure*}[ht!]
\centering
\includegraphics[width=0.9\textwidth]{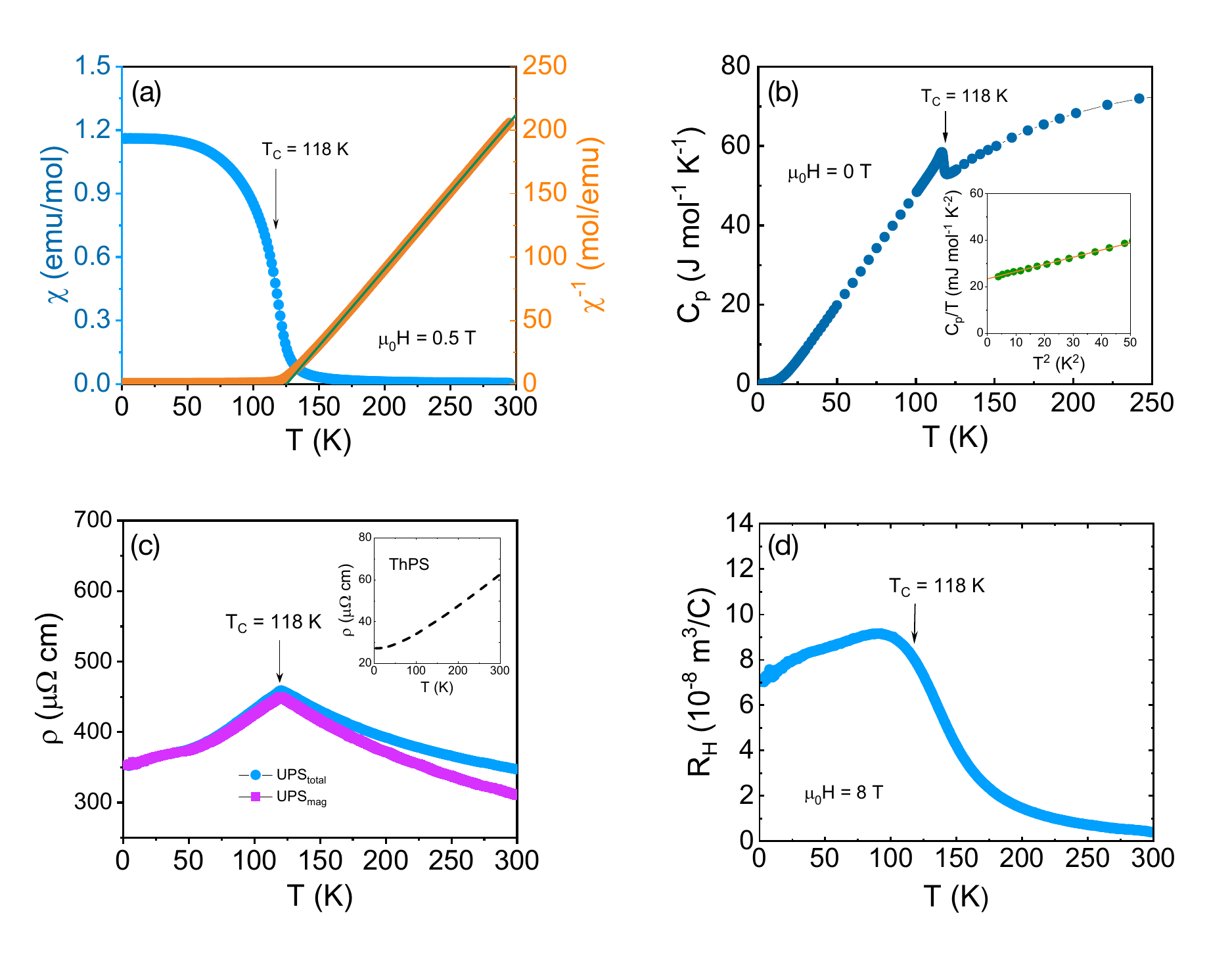}
\caption{Bulk magnetic, thermodynamic, and transport properties of UPS. (a) Magnetic susceptibility $\chi(T)$ and inverse susceptibility $\chi^{-1}(T)$ measured with $\mu_0H=0.5$~T applied along the crystallographic $c$ axis. The solid line denotes a Curie-Weiss fit in the paramagnetic regime. (b) Temperature dependence of the molar heat capacity $C_p(T)$. The inset shows $C_p/T$ as a function of $T^2$, together with a fit to $C_p/T=\gamma+\beta T^2$. (c) Longitudinal resistivity of UPS and the nonmagnetic reference compound ThPS (inset). Data for ThPS has been taken from Ref.~\cite{henkie}. The magnetic contribution (purple squares) is estimated as $\rho^{\rm UPS}_{\rm mag}(T)=\rho^{\rm UPS}_{\rm total}(T)-\rho^{\rm ThPS}_{\rm ph}(T)$. (d) Effective Hall coefficient, $R_H^{8\mathrm{T}}(T)$, measured at 8~T. Arrows mark the ferromagnetic transition at $T_C=118$~K.}
\label{fig1}
\end{figure*}

Uranium compounds are ideally suited to address this question. Among correlated quantum materials, they occupy a unique position because their $5f$ electrons reside at the boundary between localized and itinerant behavior \cite{Zwicknagl2007,Moore2009,Fujimori2016,Fulde2006,Fulde2012,ROPP2017,Regmi}. The simultaneous presence of strong Coulomb interactions, strong spin-orbit coupling, sizable $5f$-$p$ hybridization, and relatively extended $5f$ orbitals produces an exceptionally rich electronic landscape in which electronic correlations, magnetic order, and nontrivial band topology naturally coexist\cite{Moore2009,Fujimori2016,Regmi}. Unlike rare-earth Kondo systems, whose low-energy physics is largely governed by localized $4f$ moments, uranium compounds typically exhibit a dual itinerant/localized character of the $5f$ electrons, making them promising candidates for realizing correlated topological electronic phases. Recent theoretical studies have suggested that strong electronic correlations may cooperate with crystalline symmetry to generate topological semimetallic states beyond the conventional band picture, including correlated Weyl semimetals and related topological phases \cite{Regmi,Lai2018,Chen2022,Ivanov2019}. In uranium-based materials, this possibility is particularly appealing because ferromagnetic order simultaneously breaks time-reversal symmetry while strong spin-orbit coupling lifts band degeneracies, providing favorable conditions for Weyl fermions and large anomalous Hall responses. Nevertheless, direct experimental studies establishing how the temperature evolution of correlated $5f$ electronic states influences anomalous Hall transport remain scarce.

Layered tetragonal uranium pnictochalcogenides, U$XY$ ($X=$ pnictogen, $Y=$ chalcogen), provide an excellent platform for investigating these questions. They crystallize in the nonsymmorphic PbFCl-type structure (space group $P4/nmm$), the same structural framework as the ZrSiS family of topological semimetals, while introducing strong electronic correlations and intrinsic ferromagnetism through the uranium $5f$ electrons. These compounds exhibit strong Ising-like ferromagnetism, pronounced magnetocrystalline anisotropy, and signatures of correlated $5f$ physics arising from hybridization between uranium $5f$ and ligand $p$ states \cite{Schoop2016,guziewicz,Zygmunt1972,Leciejewicz1972,Kaczorowski1994,Bazan1972,Zygmunt1973,Zygmunt1974-1,Zygmunt1974-2,Blaise1980,Troc1987}. Hallmarks of electronic correlations, including enhanced Sommerfeld coefficients, logarithmic electrical resistivity, and anomalous Hall transport, have long been recognized in this family \cite{Wojakowski1972,Henkie1995,Chicorek2001,Wojakowski2001}. More recently, USbTe was identified as a correlated magnetic topological semimetal exhibiting a large anomalous Hall conductivity together with unconventional anomalous Hall scaling in the Kondo-coherent regime, highlighting the important role of electronic correlations in shaping topological transport \cite{Siddiquee2023}. Similar behavior has subsequently been reported in UBiTe \cite{Xu2024}, establishing uranium pnictochalcogenides as a broader family of correlated magnetic semimetals. Whether the evolution of anomalous Hall transport generally follows the development of magnetic order or the reconstruction of the correlated $5f$ electronic structure, however, remains an open question.

\begin{figure}[t]
\centering
\includegraphics[width=0.5\textwidth]{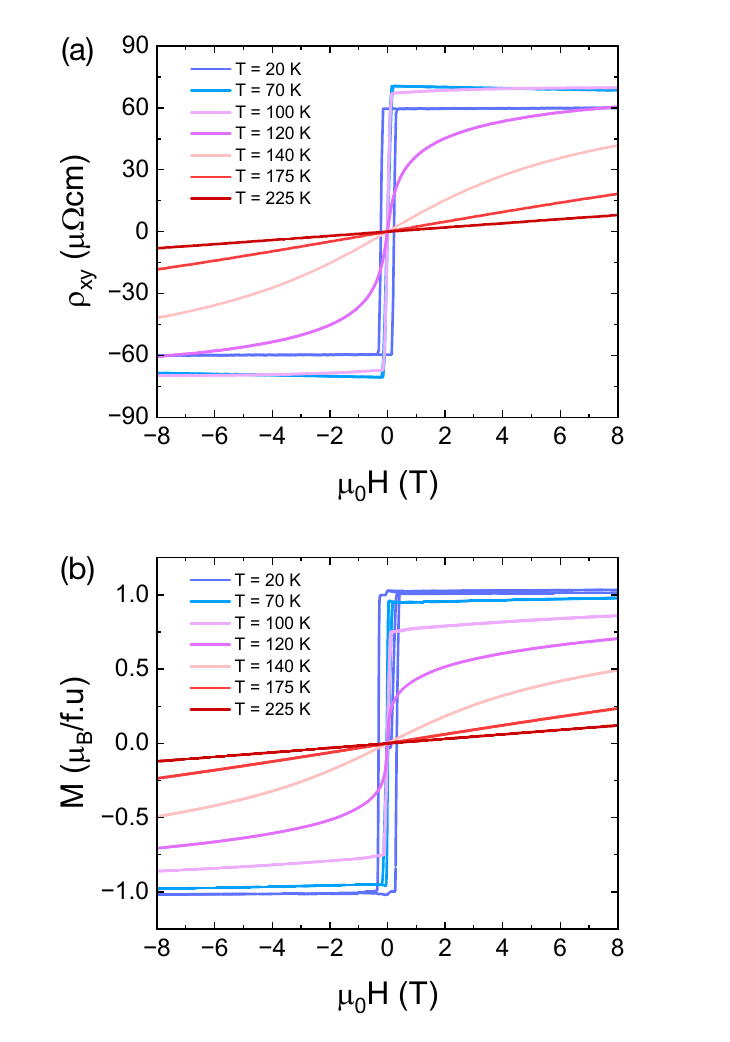}
\caption{Hall resistivity and magnetization isotherms. (a) Hall resistivity $\rho_{xy}(H)$ measured at selected temperatures. At temperatures well above $T_C$, the Hall response is approximately linear in field. A nonlinear contribution becomes apparent on approaching the transition and develops into a strongly nonlinear, hysteretic response within the ferromagnetic state. (b) Corresponding magnetization isotherms measured with the magnetic field applied along the crystallographic $c$ axis. The development of hysteresis closely follows the evolution of the Hall response, indicating its magnetic origin.}
\label{fig2}
\end{figure}

In this work, we investigate the uranium ferromagnet UPS using Hall transport, angle-resolved photoemission spectroscopy (ARPES), thermodynamic measurements, and density-functional calculations. Resonant ARPES establishes the dual nature of the uranium $5f$ electrons, revealing a narrow low-energy $5f$ feature at the Fermi level together with broad incoherent spectral weight at higher binding energies. Temperature-dependent measurements further reveal that the low-energy spectral weight changes only weakly across $T_C=118$~K but increases below approximately 90~K, demonstrating that the correlated electronic structure continues to evolve well inside the ferromagnetic state. Hall measurements reveal a large anomalous Hall conductivity approaching $4.5\times10^{2}\,\Omega^{-1}\,\mathrm{cm}^{-1}$ that develops progressively across the ferromagnetic transition and does not simply follow the ordered magnetic moment. First-principles calculations identify a symmetry-protected Weyl crossing with pronounced Berry curvature and yield a large intrinsic anomalous Hall conductivity, providing a microscopic connection between the topological electronic structure and the observed Hall response. Together, these results establish UPS as a correlated uranium Weyl semimetal in which the anomalous Hall response develops while the correlated uranium $5f$ electronic structure is itself still evolving.

\textit{Methods} - Single crystals of UPS were grown by the chemical vapor transport method using iodine as the transport agent and structurally characterized by x-ray diffraction, confirming the previously reported tetragonal PbFCl-type crystal structure (space group $P4/nmm$), in which uranium atoms form square nets separated by alternating P and S layers along the crystallographic $c$ axis \cite{Kaczorowski1994}. Magnetization, heat capacity, electrical resistivity, and Hall-effect measurements were performed using the vibrating sample magnetometer, heat-capacity, and resistivity options of a Quantum Design DynaCool Physical Property Measurement System (PPMS). Hall measurements were carried out with the magnetic field applied along the crystallographic $c$ axis and current flowing within the basal plane. ARPES measurements were performed at beamlines 10.0.1.2 and 4.0.3.2 of the Advanced Light Source (ALS), Lawrence Berkeley National Laboratory. Electronic structure calculations were performed within density functional theory including spin-orbit coupling, while the Berry curvature and intrinsic anomalous Hall conductivity were evaluated using a Wannier-interpolated Hamiltonian. Additional experimental and computational details are provided in the Supplemental Material \cite{SM}.

\begin{figure*}[t]
\centering
\includegraphics[width=1\textwidth]{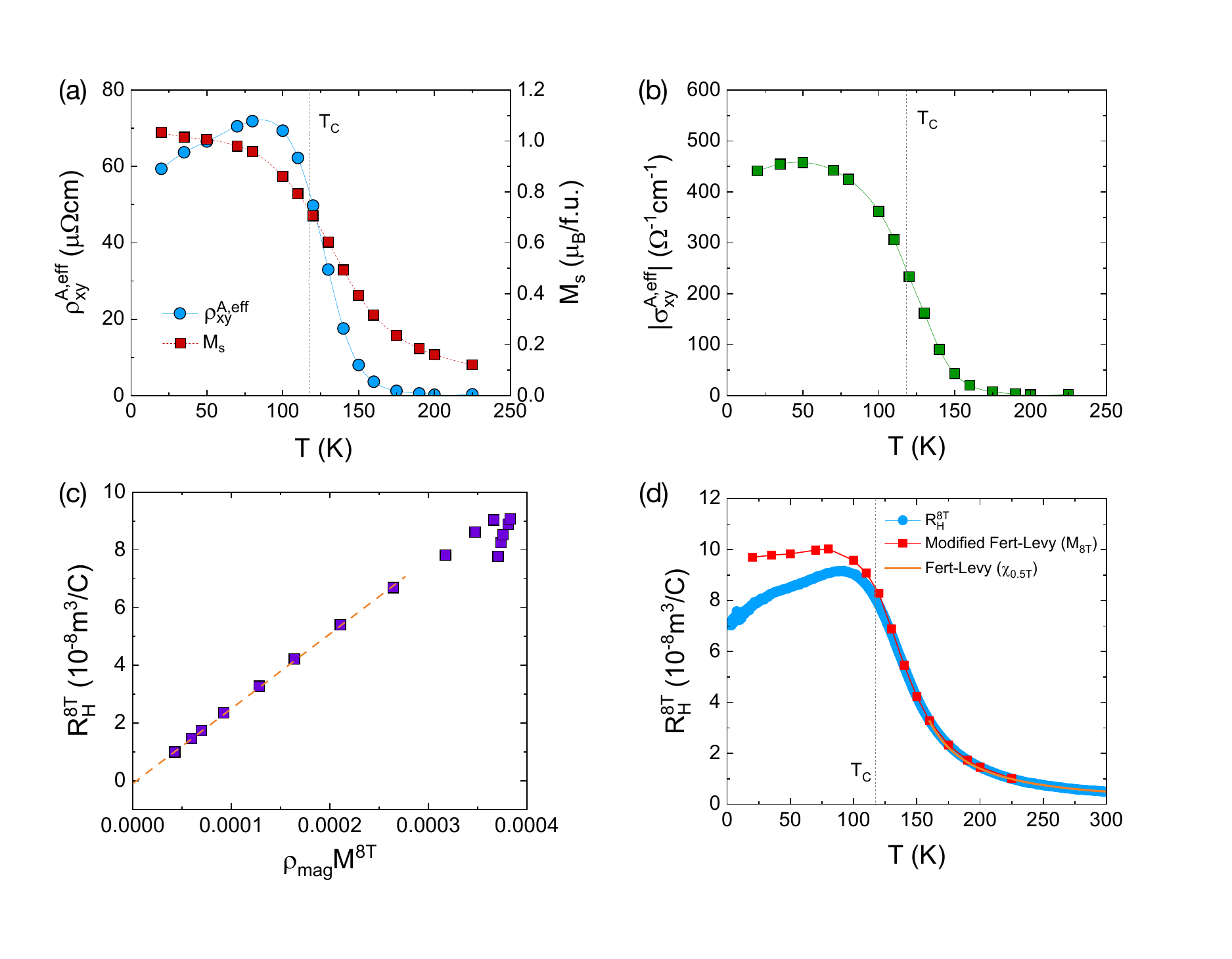}
\caption{Temperature evolution and scaling analysis of the anomalous Hall response in UPS. (a) Effective anomalous Hall resistivity, $\rho_{xy}^{A,\mathrm{eff}}$, compared with the saturation magnetization $M_s$. While both quantities increase below $T_C$, the anomalous Hall resistivity exhibits a broad maximum near 80-90~K and does not simply follow the ordered magnetic moment. (b) Effective anomalous Hall conductivity, $|\sigma_{xy}^{A,\mathrm{eff}}|$, showing the rapid development of a large anomalous Hall response below the Curie temperature. (c) Modified Fert-Levy scaling of the high-temperature Hall coefficient, $R_H^{8\mathrm{T}}$, as a function of $\rho_{\mathrm{mag}}M_{8\mathrm{T}}$. The solid line is a linear fit to the high-temperature data. (d) Comparison of the measured Hall coefficient with calculations based on the modified Fert-Levy model and the conventional susceptibility-based formulation. The modified scaling reproduces the Hall response in the paramagnetic state but progressively deviates below $T_C$, indicating that magnetic skew scattering alone cannot fully account for the low-temperature Hall response.}
\label{fig3}
\end{figure*}

\textit{Bulk properties} - The bulk magnetic, thermodynamic, and transport properties of UPS are summarized in Fig.~\ref{fig1}. The magnetic susceptibility measured with an applied field of $\mu_0H=0.5$~T parallel to the $c$ axis is shown in Fig.~\ref{fig1}(a). A pronounced increase in $\chi(T)$ marks the onset of ferromagnetic order at $T_C=118$~K, consistent with previous reports \cite{Kaczorowski1994}. Above the ordering temperature, the susceptibility follows the Curie-Weiss law. A fit to the paramagnetic region yields an effective magnetic moment of $\mu_{\rm eff}=2.57~\mu_B$/U and a Curie-Weiss temperature $\theta_P=125$~K, indicating dominant ferromagnetic exchange interactions. At 20~K, the magnetization reaches $M_s=1.03~\mu_B$/f.u. (see Fig.~\ref{fig2} and the Supplemental Material \cite{SM}), substantially smaller than the free-ion LS-coupling saturation moments of $3.27~\mu_B$ for U$^{3+}$ and $3.20~\mu_B$ for U$^{4+}$. Although crystal-field effects and incomplete field-induced saturation may also contribute, the strongly reduced moment is consistent with substantial $5f$ hybridization and the partially itinerant character of the uranium $5f$ electrons \cite{Zwicknagl2007,Fujimori2016,Durakiewicz2014}. The heat capacity shown in Fig.~\ref{fig1}(b) exhibits a well-defined $\lambda$-type anomaly at $T_C$, confirming the bulk nature of the ferromagnetic transition. Fitting the low-temperature data using $C_p(T)=\gamma T+\beta T^3$, yields an electronic specific-heat coefficient $\gamma=23$~mJ\,mol$^{-1}$\,K$^{-2}$ and a phonon coefficient $\beta=0.309$~mJ\,mol$^{-1}$\,K$^{-4}$, corresponding to a Debye temperature $\Theta_D=267$~K (see also Ref.~\cite{wojakowski}). The longitudinal resistivity is presented in Fig.~\ref{fig1}(c), together with the nonmagnetic reference compound ThPS taken from Ref.~\cite{henkie}. The magnetic contribution, estimated as $\rho^{\rm UPS}_{\rm mag}(T)=\rho^{\rm UPS}_{\rm total}(T)-\rho^{\rm ThPS}_{\rm ph}(T)$, exhibits an extended $-\ln T$ dependence that persists nearly to $T_C$, reflecting strong spin-dependent scattering associated with correlated uranium $5f$ electrons. In contrast to many conventional Kondo-lattice compounds, UPS does not exhibit a pronounced coherence maximum above the magnetic transition, suggesting that the low-energy electronic structure continues to evolve within the ferromagnetic state. The corresponding Hall coefficient, $R_H^{8{\rm T}}(T)$ obtained at 8~T, is shown in Fig.~\ref{fig1}(d). Upon cooling below approximately 150~K, $R_H^{8{\rm T}}(T)$ increases rapidly, reaches a broad maximum near 80~K, and decreases again at lower temperatures. The broad evolution already suggests that the Hall response cannot be described by the magnetic order parameter alone, motivating the detailed Hall analysis presented below.

Figure~\ref{fig2} presents representative Hall resistivity and magnetization isotherms measured across the ferromagnetic transition. The complete sets of $\rho_{xy}(H)$ and $M(H)$ measurements are provided in the Supplemental Material \cite{SM}. At high temperatures, the Hall resistivity is approximately linear in magnetic field, consistent with an ordinary Hall response modified by magnetic scattering. Upon cooling, a nonlinear contribution becomes increasingly apparent as the system approaches the ferromagnetic transition. Below $T_C$, the Hall signal develops pronounced nonlinearity and hysteresis that closely follow the evolution of the magnetization. At the lowest temperatures, the Hall resistivity reaches approximately $70~\mu\Omega$\,cm at high field. Although the field dependence of $\rho_{xy}(H)$ closely resembles that of the magnetization, their temperature evolutions differ substantially.

Figure~\ref{fig3}(a) compares the effective anomalous Hall resistivity, $\rho_{xy}^{A,\mathrm{eff}}$, with the saturation magnetization. For each temperature, the Hall resistivity was fitted over the
high-field interval $6 \leq \mu_0H \leq 8$~T using
\begin{equation}
\rho_{xy}(\mu_0H)=R_H^{\mathrm{HF}}\mu_0H+\rho_{xy}^{A,\mathrm{eff}},
\end{equation}
where $R_H^{\mathrm{HF}}$ is the high-field Hall slope and $\rho_{xy}^{A,\mathrm{eff}}$ is the zero-field intercept of the
linear fit. The fitting parameters were determined independently at each temperature. Both $\rho_{xy}^{A,\mathrm{eff}}$ and $M_s$ increase rapidly below $T_C$, reflecting their common magnetic origin. However, while the ordered moment increases monotonically on cooling, $\rho_{xy}^{A,\mathrm{eff}}$ exhibits a broad maximum near 90-100~K before decreasing toward low temperatures. Consequently, the anomalous Hall response cannot be described simply by the temperature dependence of the ordered magnetic moment.

Using the measured longitudinal resistivity, the effective anomalous Hall conductivity was calculated according to

\begin{equation}
\sigma_{xy}^{A,\mathrm{eff}}=-\frac{\rho_{xy}^{A,\mathrm{eff}}}{\rho_{xx}^{2}+\rho_{xy}^{2}}.
\end{equation}

The resulting temperature dependence is presented in Fig.~\ref{fig3}(b). Below $T_C$, $|\sigma_{xy}^{A,\mathrm{eff}}|$ increases rapidly and reaches approximately $4.5\times10^{2}~\Omega^{-1}\mathrm{cm}^{-1}$ at low temperature, comparable to the values recently reported for the uranium magnetic semimetals USbTe and UBiTe \cite{Siddiquee2023,Xu2024}. Although the effective anomalous Hall resistivity exhibits a broad maximum near 90~K [Fig.~\ref{fig3}(a)], the corresponding anomalous Hall conductivity
tends to saturate at low temperatures because the decrease in $\rho_{xy}^{A,\mathrm{eff}}$ is largely compensated by the continued decrease of the longitudinal resistivity [Eq.~(2)]. The large magnitude of the anomalous Hall conductivity suggests a substantial intrinsic contribution, a conclusion supported below by the calculated Berry curvature and intrinsic anomalous Hall conductivity.

To examine whether the Hall response can be understood within conventional magnetic skew-scattering theory, we analyze the high-temperature Hall coefficient using a modified Fert-Levy scaling relation,
\begin{equation}
R_H^{8\mathrm{T}}=R_0+A_M\rho_{\mathrm{mag}}M_{8\mathrm{T}},
\end{equation}
where $M_{8\mathrm{T}}$ denotes the magnetization measured at the same magnetic field as the Hall effect. Unlike the original Fert-Levy formulation \cite{FL1,FL2}, which employs the low-field susceptibility, this expression avoids complications arising from the strong field dependence of the magnetization below the ferromagnetic transition. The fit to the high-temperature data was performed for $T\ge150$~K, yielding $R_0=-1.19\times10^{-9}\ {\rm m^3C^{-1}}$ and $A_M=2.646\times10^{-4}$. 

The scaling is illustrated in Fig.~\ref{fig3}(c,d). In the paramagnetic state, the measured Hall coefficient follows the modified Fert-Levy relation remarkably well, indicating that incoherent magnetic skew scattering dominates the Hall response. On entering the ferromagnetic phase, however, the measured Hall coefficient progressively deviates from this behavior. The discrepancy is substantially smaller than obtained using the conventional susceptibility-based formulation, yet a systematic low-temperature deviation remains. This observation indicates that magnetic skew scattering alone cannot fully account for the Hall response once long-range ferromagnetic order is established. These results suggest that the evolution of the Hall response is governed not only by magnetic scattering but also by changes in the underlying electronic structure. To investigate this possibility, we examine the low-energy uranium $5f$ states using resonant ARPES. 

\begin{figure}[b]
\centering
\includegraphics[width=0.5\textwidth]{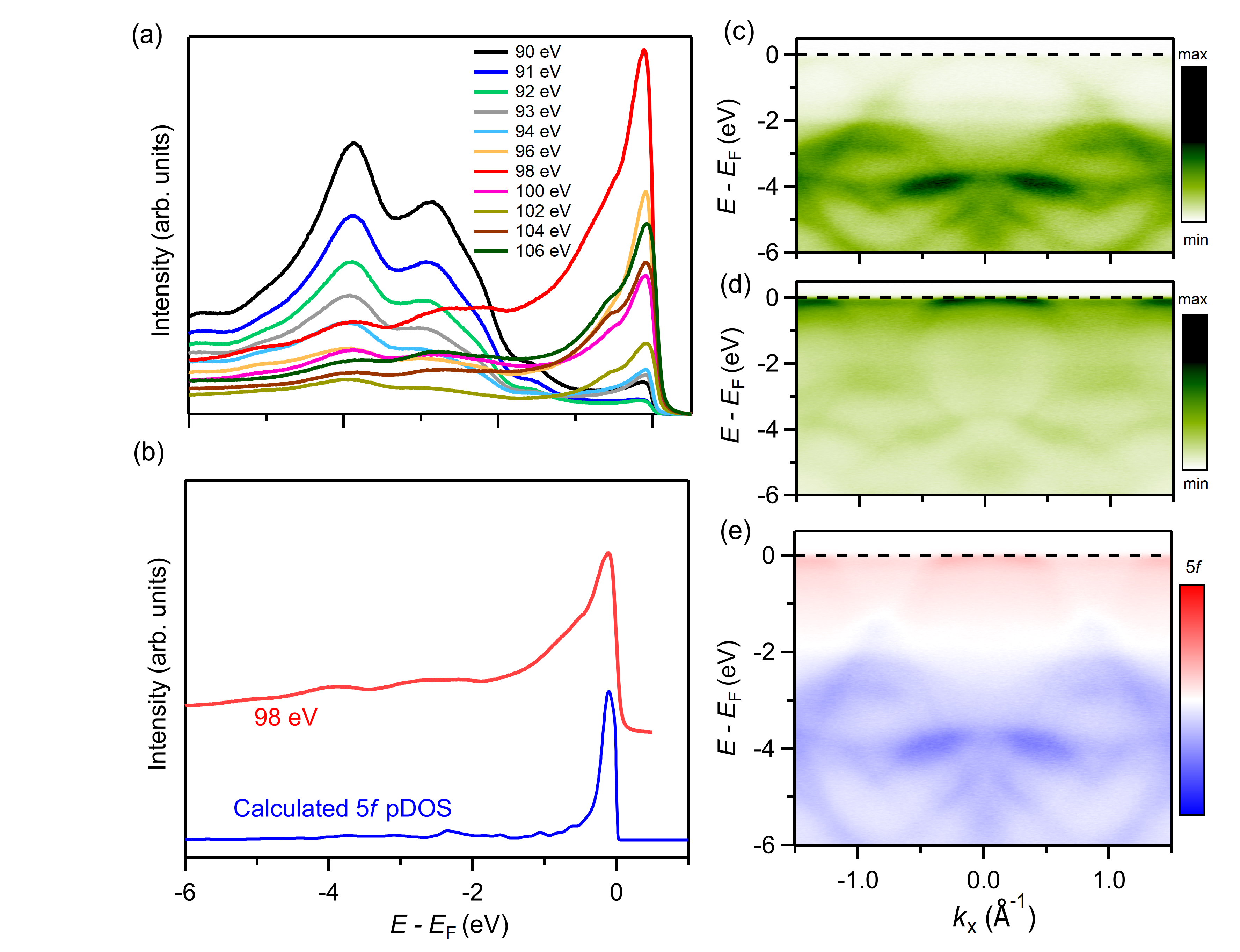}
\caption{Resonant enhancement of the uranium $5f$ electronic states, demonstrating the dual itinerant-correlated nature of the uranium $5f$ electrons. (a) Angle-integrated valence-band spectra measured for photon energies between 90 and 106~eV across the U $5d$-$5f$ absorption edge. (b) Valence-band spectrum measured at the resonant photon energy ($h\nu=98$~eV, red) compared with the calculated U-$5f$ partial density of states (blue), multiplied by the Fermi-Dirac distribution. (c,d) Wide-energy momentum-resolved spectra measured under off-resonant (92~eV) and resonant (98~eV) conditions along the $\overline{X}-\overline{\Gamma}-\overline{X}$ direction. (e) Normalized difference between the resonant and off-resonant spectra, highlighting the uranium $5f$ spectral weight selectively enhanced under resonant conditions. Data were collected at the ALS beamline 10.0.1.2 at $T=15$~K.}
\label{fig4}
\end{figure}

\begin{figure*}[t]
\centering
\includegraphics[width=0.90\textwidth]{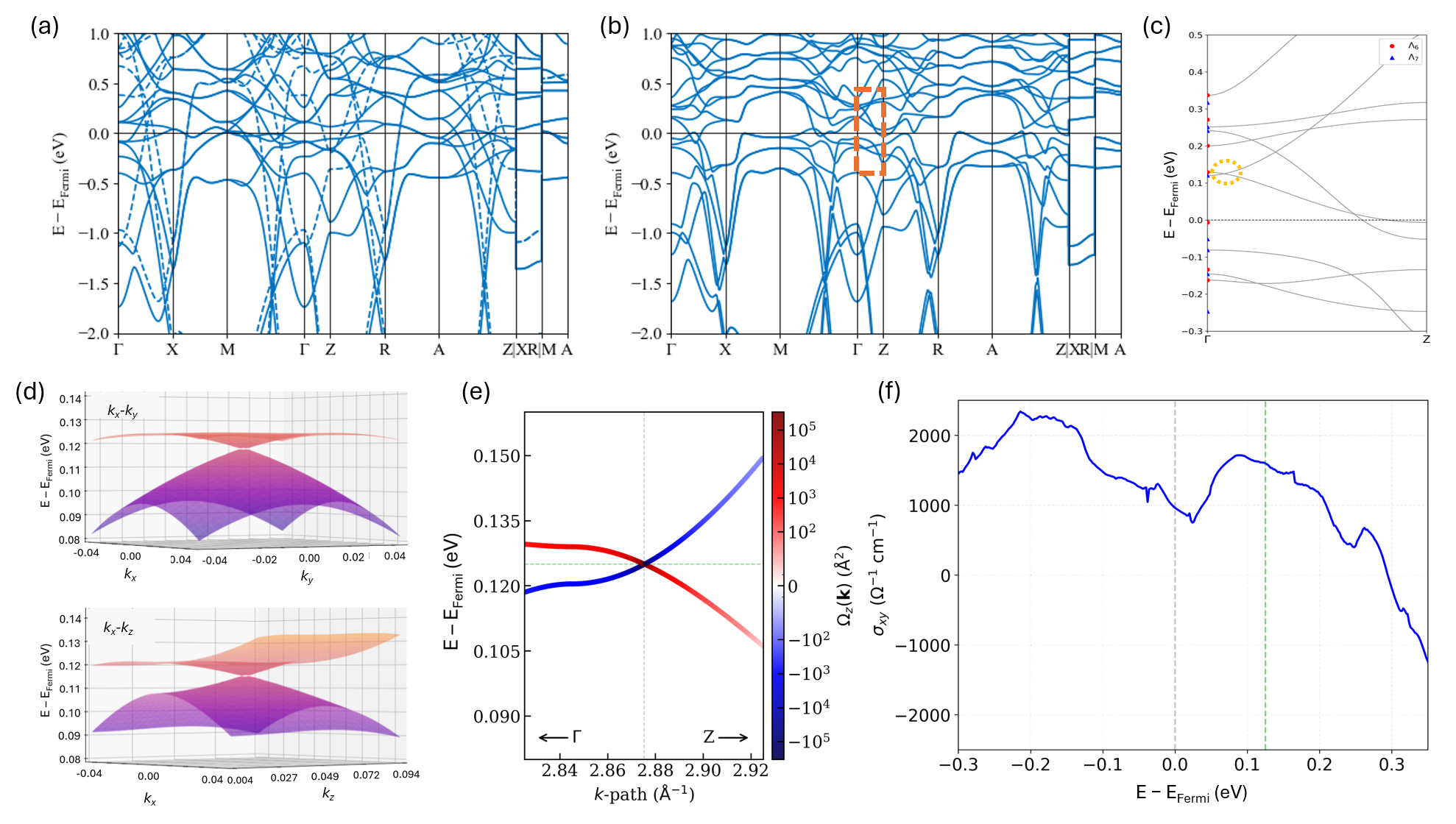}
\caption{Density-functional calculations of the bulk electronic structure of ferromagnetic UPS. (a,b) Calculated bulk band structure without (a) and with (b) spin-orbit coupling. The solid and dashed bands in (a) represent spin up and down, respectively. (c) Enlarged view of the band structure along the $\Gamma-Z$ direction within the dashed rectangle in (b). The dashed circle highlights a symmetry-protected band crossing between bands belonging to different irreducible representations. (d) Calculated dispersions in the $k_x$--$k_y$ (top) and $k_x$--$k_z$ (bottom) planes through the crossing, demonstrating its three-dimensional linear dispersion characteristic of a Weyl point. (e) Berry-curvature-projected bands in the vicinity of the Weyl crossing along the $\Gamma-Z$ direction, showing the pronounced sign change of $\Omega_z(\mathbf{k})$ across the crossing. (f) Calculated intrinsic anomalous Hall conductivity, $\sigma_{xy}^{\mathrm{int}}$, as a function of energy relative to the Fermi level. The gray and green dashed lines mark $E_F$ and the Weyl-point energy, respectively.}
\label{fig5}
\end{figure*}

\textit{Electronic properties} - Figure~\ref{fig4}(a) presents angle-integrated valence-band spectra measured for photon energies between 90 and 106~eV across the U $5d$-$5f$ absorption edge and along the $\overline{X}-\overline{\Gamma}-\overline{X}$ direction. At photon energies below the U $5d$-$5f$ absorption edge (90-94~eV), the spectra are dominated by broad valence-band features extending several electron volts below the Fermi level. As the photon energy approaches the resonance, a sharp feature develops at the Fermi level and reaches maximum intensity at $h\nu=98$~eV, corresponding to the U $5d$-$5f$ absorption edge. The low-energy spectral weight is strongly suppressed near 92~eV, consistent with measurements close to the anti-resonance condition.

The resonantly enhanced spectrum is compared with the calculated U-$5f$ partial density of states in Fig.~\ref{fig4}(b). The sharp feature at the Fermi level closely follows the calculated U-$5f$ density of states, demonstrating that the low-energy electronic states have predominantly uranium $5f$ character. In contrast, a broad spectral hump extending beyond 2~eV binding energy remains visible under resonant conditions, indicating that a substantial fraction of the uranium $5f$ spectral weight resides in broad, weakly dispersive excitations. The coexistence of a narrow low-energy $5f$ feature with a broad spectral weight at higher binding energies is characteristic of the dual itinerant-correlated nature of the uranium $5f$ electrons. Figures~\ref{fig4}(c-e) compare wide-energy momentum-resolved spectra measured under off-resonant ($h\nu=92$~eV) and resonant ($h\nu=98$~eV) conditions. Resonant excitation selectively enhances the spectral weight within approximately 0.5~eV of the Fermi level while leaving the deeper valence bands comparatively unchanged. To visualize this enhancement, Fig.~\ref{fig4}(e) presents the normalized difference between the resonant and off-resonant spectra after scaling each map to its respective maximum intensity. The resonantly enhanced intensity is concentrated near the Fermi level and follows the dispersive low-energy bands, confirming their strong uranium $5f$ character. Its pronounced momentum dependence further demonstrates that the uranium $5f$ electrons participate directly in the itinerant low-energy electronic structure.

\begin{figure*}[t]
\centering
\includegraphics[width=0.9\textwidth]{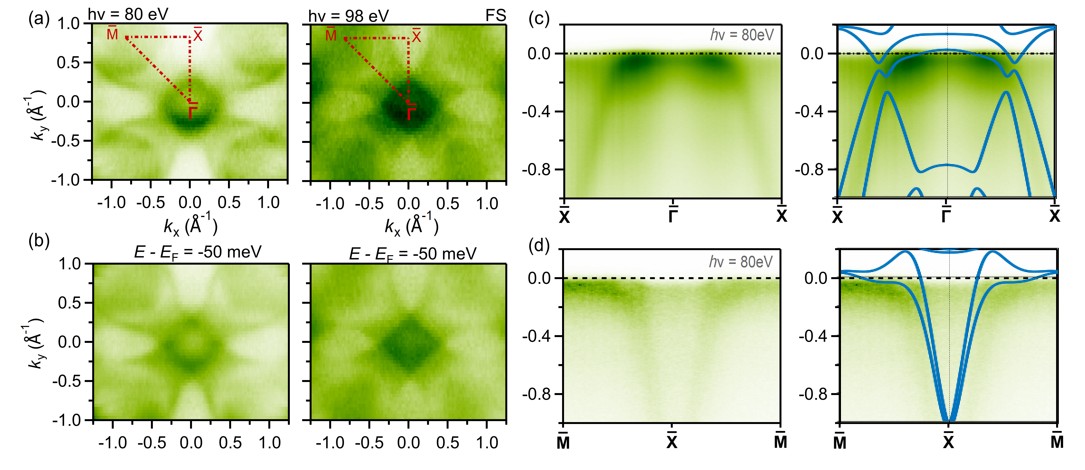}
\caption{Comparison of the experimental and calculated electronic structure of UPS. (a,b) Fermi surface and constant-energy contour measured 50~meV below the Fermi level using off-resonant ($h\nu=80$~eV, left) and resonant ($h\nu=98$~eV, right) photon energies. Resonant excitation selectively enhances the uranium $5f$ spectral weight near the $\overline{\Gamma}$ and $\overline{M}$ points. (c,d) Experimental ARPES intensity maps (left) and calculated bulk band structure (right) along the $\overline{X}-\overline{\Gamma}-\overline{X}$ and $\overline{M}-\overline{X}-\overline{M}$ directions, respectively. While the calculations reproduce the overall dispersive band structure, noticeable differences remain near the Fermi level, reflecting many-body renormalization of the low-energy uranium $5f$ electronic states. Data were collected at the ALS beamline 10.0.1.2 at $T=15$~K.}
    \label{fig6}
\end{figure*}

\begin{figure*}[t]
\centering
\includegraphics[width=0.8\textwidth]{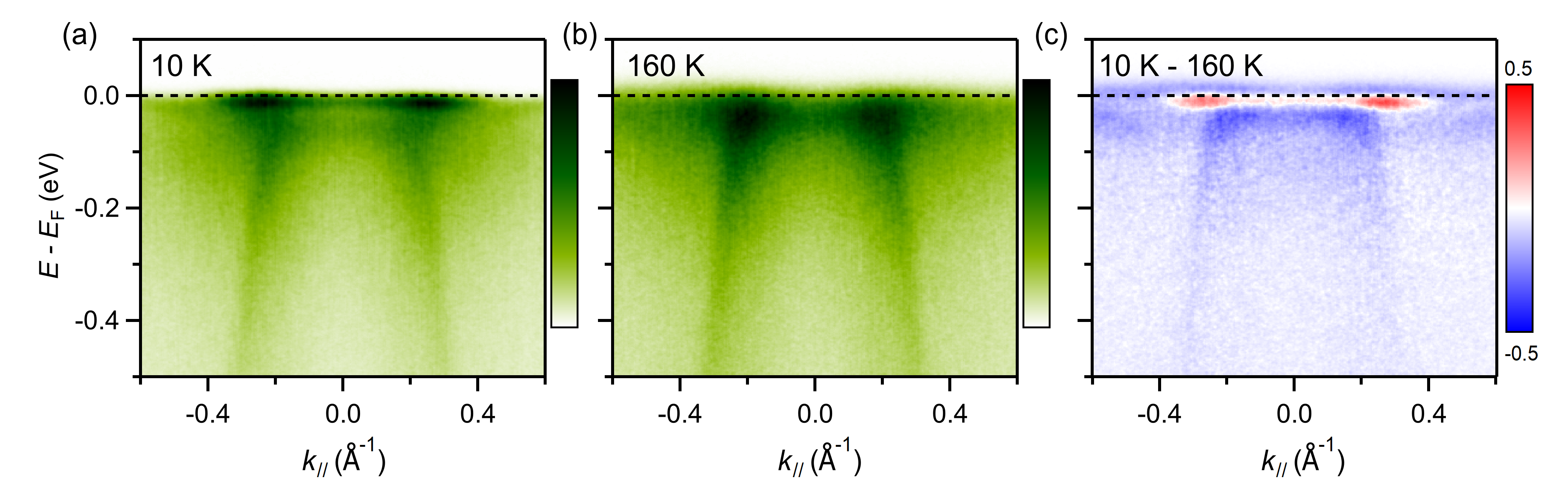}
\caption{Temperature evolution of the low-energy electronic structure across the ferromagnetic transition.
(a,b) ARPES intensity maps measured along the $\overline{M}-\overline{\Gamma}-\overline{M}$ direction at 10~K (ferromagnetic state) and 160~K (paramagnetic state), respectively. (c) Normalized difference map obtained after scaling both spectra to their respective maximum intensities before subtraction. The largest temperature-induced changes are concentrated in the vicinity of the Fermi level, whereas the deeper valence bands remain largely unchanged. Data were collected at the ALS beamline 4.0.3.2 using a photon energy of $h\nu=80$~eV.}
\label{fig7}
\end{figure*}

Figure~\ref{fig5} presents the calculated bulk electronic structure of ferromagnetic UPS within density functional theory including spin-orbit coupling. The electronic states in the vicinity of the Fermi level are dominated by uranium $5f$ orbitals, consistent with the resonant ARPES measurements. Fig.~\ref{fig5}(a) shows the band structure without spin-orbit coupling (SOC), where spin-up and spin-down channels are plotted as solid and dashed lines, respectively. When SOC is included, these channels are mixed, and thus plotted using uniform solid lines in Fig.~\ref{fig5}(b). A prominent feature of the calculated band structure is the symmetry-protected crossing along the $\Gamma-Z$ direction, highlighted by the dashed circle in Fig.~\ref{fig5}(c). The crossing occurs between bands belonging to different irreducible representations of the $C_4$ rotational symmetry and therefore remains protected in the presence of spin-orbit coupling. The calculated dispersions in the $k_x-k_y$ and $k_x-k_z$ planes, shown in Fig.~\ref{fig5}(d), demonstrate that the crossing is linear along all three momentum directions, identifying it as a Weyl point rather than a nodal-line crossing. The Weyl node appears approximately 0.13~eV above the Fermi level and is a direct consequence of the ferromagnetic ground state. The precise energy of the calculated crossing should, however, be
interpreted with caution because the low-energy uranium $5f$ states exhibit substantial many-body renormalization beyond the present density-functional treatment. Ferromagnetic order breaks time-reversal symmetry, lifts the Kramers degeneracy, and permits isolated crossings between nondegenerate bands. Such magnetic Weyl nodes are expected to generate strong Berry curvature and therefore provide a natural microscopic framework for understanding the large anomalous Hall response observed experimentally \cite{Nagaosa2010,Armitage2018,Siddiquee2023}. The Berry-curvature-projected band structure in Fig.~\ref{fig5}(e) reveals a pronounced sign-changing Berry-curvature distribution associated with the symmetry-protected Weyl crossing along $\Gamma$-$Z$. Integration of the Berry curvature over the Brillouin zone yields the intrinsic anomalous Hall conductivity shown in Fig.~\ref{fig5}(f). At the calculated Fermi level, $\sigma_{xy}^{\mathrm{int}}\simeq9.6\times10^{2}~\Omega^{-1}\mathrm{cm}^{-1}$, of the same order of magnitude as the experimental low-temperature value of approximately $4.5\times10^{2}~\Omega^{-1}\mathrm{cm}^{-1}$. Together, these calculations support a substantial intrinsic contribution to the anomalous Hall response in UPS.

The calculated band structure provides the framework for interpreting the momentum-resolved ARPES measurements shown in Fig.~\ref{fig6}. Comparison between experiment and theory allows the role of electronic correlations in the low-energy uranium $5f$ electronic structure to be assessed directly. Figures~\ref{fig6}(a,b) present the Fermi surface and constant-energy contours measured using off-resonant (80~eV) and resonant (98~eV) photon energies. Resonant excitation produces a pronounced enhancement of the spectral weight around the $\overline{\Gamma}$ and $\overline{M}$ points and throughout the low-energy constant-energy contours, again demonstrating that these states possess strong uranium $5f$ character.

\begin{figure*}[t]
\centering
\includegraphics[width=1\textwidth]{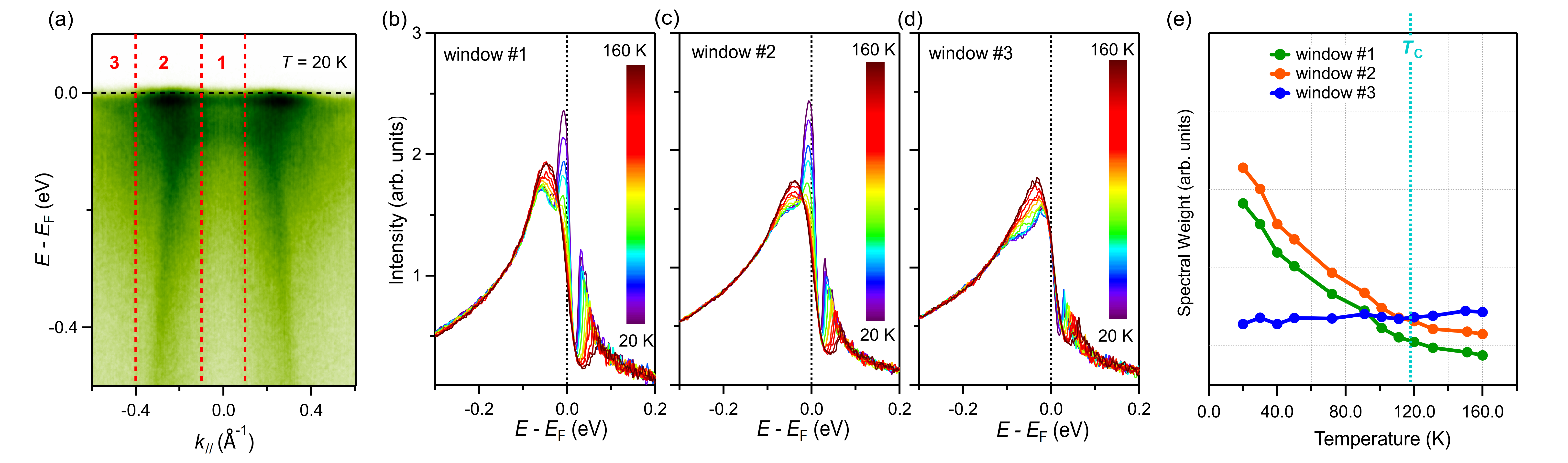}
\caption{Temperature evolution of the low-energy uranium $5f$ electronic structure.
(a) ARPES intensity map measured along the $\overline{M}-\overline{\Gamma}-\overline{M}$ direction. The dashed lines indicate the momentum windows used for extracting energy distribution curves (EDCs). (b-d) Temperature evolution of the EDCs, divided by the Fermi-Dirac distribution, for windows~1-3 defined in (a). The spectra are normalized to the intensity within the $-0.18$ to $-0.12$~eV energy interval. (e) Integrated spectral weight within $E_F\pm10$~meV for the three momentum windows. Windows 1 (green) and 2 (orange) show relatively little change across $T_C$, followed by a pronounced increase in low-energy spectral weight below approximately 90~K, demonstrating that the reconstruction of the low-energy uranium $5f$ electronic structure develops well inside the ferromagnetic state.}
\label{fig8}
\end{figure*}

The corresponding band dispersions measured along the $\overline{X}-\overline{\Gamma}-\overline{X}$ and $\overline{M}-\overline{X}-\overline{M}$ directions are compared with the calculated bulk bands in Figs.~\ref{fig6}(c,d). The overall agreement is good for the strongly dispersive valence bands extending beyond 1~eV binding energy. In contrast, the experimental spectra exhibit substantially stronger low-energy spectral weight near the $\overline{\Gamma}$ point than predicted by density-functional theory. These discrepancies originate from electronic correlation effects that are not fully captured within the static mean-field approximation. While density-functional theory successfully reproduces the overall band topology, the low-energy uranium $5f$ states undergo substantial many-body renormalization. The remaining question is whether this renormalization directly tracks the magnetic transition or continues to evolve within the ordered state.

To investigate the evolution of the low-energy electronic structure across the ferromagnetic transition, we compare ARPES spectra collected in the ferromagnetic (10 K) and paramagnetic (160 K) states. Figures~\ref{fig7}(a,b) show the band dispersion measured along the $\overline{M}-\overline{\Gamma}-\overline{M}$ direction. At low temperature, a shallow electron-like band, a strongly dispersive hole-like band extending beyond 1~eV binding energy, and a nearly dispersionless feature at the Fermi level are clearly resolved. The corresponding difference map, obtained after normalizing both spectra to their maximum intensity [Fig.~\ref{fig7}(c)], demonstrates that the largest temperature-induced changes are confined to the vicinity of the Fermi level. In particular, the low-temperature spectrum exhibits enhanced spectral weight associated with the flat $5f$-derived feature, while the deeper valence bands remain largely unchanged.

To further track this evolution, we examine ARPES spectra acquired at intermediate temperatures (see Supplemental Material~\cite{SM}). The nearly dispersionless feature becomes progressively more pronounced upon
cooling. EDCs integrated within $\pm0.1$~\AA$^{-1}$ of $\overline{\Gamma}$ reveal that a corresponding peak approximately 10~meV below the Fermi level is clearly resolved by 72~K and strengthens upon further cooling, providing direct spectroscopic evidence for the development of the low-energy electronic reconstruction within the ferromagnetic state. To quantify these changes, we analyze the temperature evolution of EDCs extracted from the three momentum windows indicated in Fig.~\ref{fig8}(a). Window~1 is centered on the nearly dispersionless feature around the $\overline{\Gamma}$ point, window~2 samples the adjacent momentum region where the strongest temperature dependence is observed, and window~3 probes the outer dispersive states. The EDCs were normalized to their intensity within the energy interval between $-0.18$ and $-0.12$~eV, where the spectral weight is nearly temperature independent, and subsequently divided by the Fermi-Dirac distribution to facilitate comparison of the low-energy electronic states.

The temperature evolution of the EDCs is presented in Figs.~\ref{fig8}(b-d). Windows~1 and~2 exhibit a pronounced enhancement of spectral weight within several tens of meV of the Fermi level upon cooling. The enhancement develops continuously over an extended temperature interval rather than exhibiting a distinct anomaly at the Curie temperature, and is accompanied by a substantial redistribution of low-energy spectral weight over an energy scale of approximately 100~meV. In contrast, the spectra extracted from window~3 evolve only weakly with temperature, demonstrating that the electronic reconstruction is strongly momentum dependent.

To quantify the evolution of the low-energy electronic structure, we integrate the spectral intensity within $\pm10$~meV of the Fermi level for each momentum window. The resulting normalized spectral weights are shown in Fig.~\ref{fig8}(e). The strongest temperature dependence is observed in windows~1 and~2. The low-energy spectral weight changes only weakly on cooling through $T_C=118$~K, but increases more rapidly below approximately 90~K. In contrast, the outer momentum window exhibits only modest changes throughout the measured temperature range. The absence of a corresponding anomaly in the ARPES spectral weight at $T_C$ shows that the low-energy electronic reconstruction does not simply track the development of long range ferromagnetic order. Instead, the pronounced redistribution of uranium $5f$ spectral weight develops well inside the ferromagnetic state, revealing distinct temperature scales for magnetic ordering and the evolution of the correlated low-energy electronic structure. Remarkably, the onset of this pronounced low-energy spectral redistribution occurs within the same temperature range in which the anomalous Hall response continues to evolve strongly, establishing a close correspondence between the Hall anomaly and the reconstruction of the correlated uranium $5f$ electronic structure. The distinct temperature dependencies demonstrate that magnetic order and the correlated low-energy electronic reconstruction are closely coupled but do not simply track one another.

Finally, we note that preliminary scanning tunneling microscopy and spectroscopy (STM/S) measurements performed at $T=11$~K (see Supplemental Material~\cite{SM}) provide independent evidence for unusual low-energy electronic states at the cleaved UPS surface. Although the microscopic origin of the observed zero-bias conductance feature remains to be established, the STM/S results are consistent with the presence of a complex low-energy electronic structure inferred from the ARPES measurements.

\begin{figure*}[t]
\centering
\includegraphics[width=0.85\textwidth]{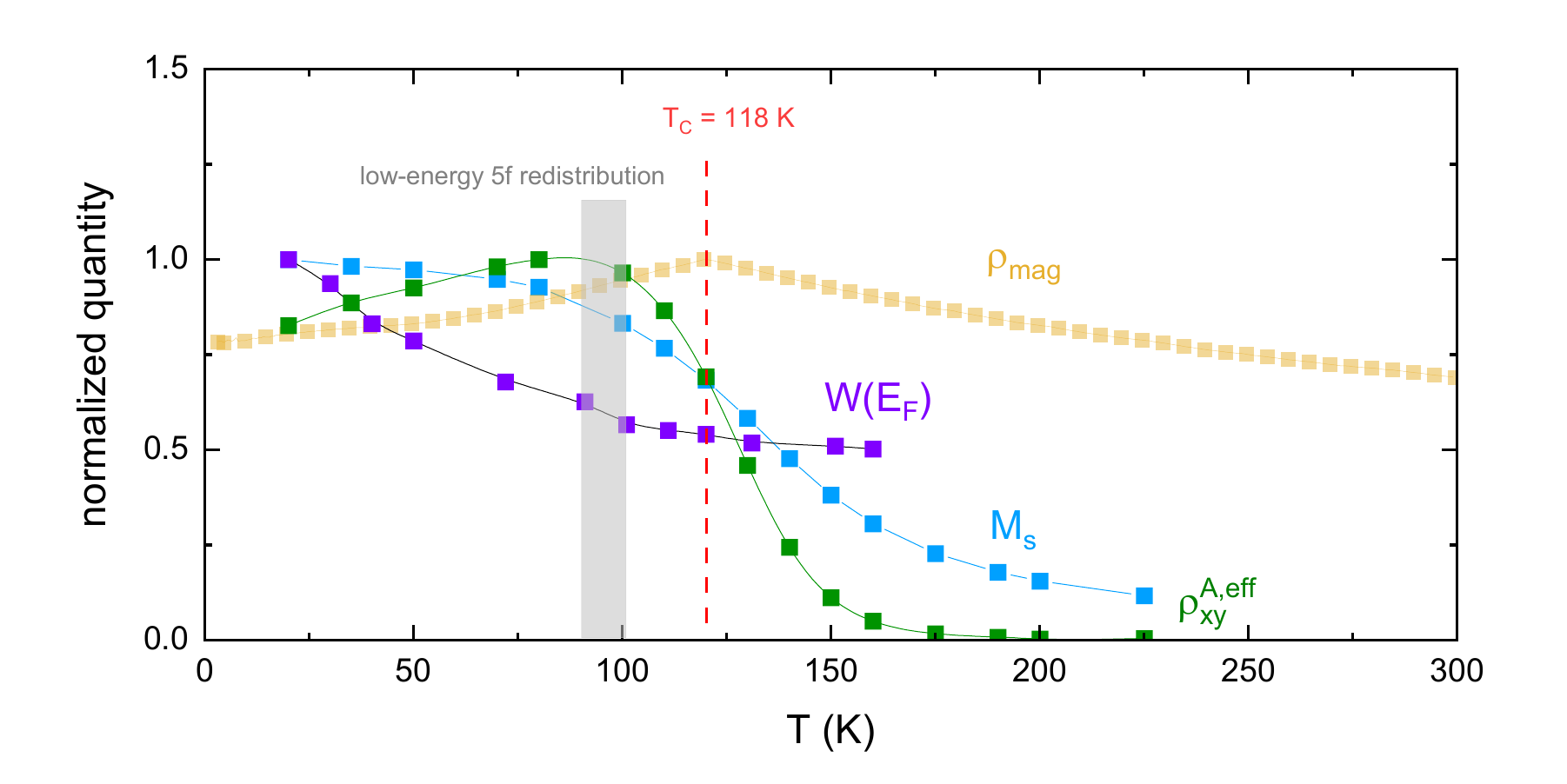}
\caption{Staged emergence of anomalous Hall transport and correlated $5f$ electronic states. Comparison of the magnetic scattering $\rho_{\mathrm{mag}}$, saturation magnetization $M_s$, effective anomalous Hall resistivity $\rho_{xy}^{A,\mathrm{eff}}$, and low-energy ARPES spectral weight $W(E_F)$, each normalized to its respective maximum value. The comparison demonstrates that magnetic scattering, long-range ferromagnetic order, anomalous Hall transport, and the redistribution of low-energy uranium $5f$ spectral weight evolve over distinct but overlapping temperature ranges. The shaded region marks the temperature interval below which ARPES reveals the pronounced redistribution of low-energy uranium $5f$ spectral weight, well inside the ferromagnetic state and within the temperature range over which the anomalous Hall response continues to evolve.}
\label{fig9}
\end{figure*}

\textit{Discussion} - The combined transport, spectroscopic, and electronic structure measurements demonstrate that the evolution of the correlated electronic structure in UPS cannot be described by a single characteristic temperature. Instead, several experimentally distinguishable energy scales emerge. Long-range ferromagnetic order develops at $T_C=118$~K, whereas temperature-dependent ARPES reveals that the pronounced redistribution of low-energy uranium $5f$ spectral weight develops only below approximately 90~K.

Figure~\ref{fig9} summarizes this behavior by comparing the normalized magnetic scattering $\rho_{\rm mag}$, saturation magnetization $M_s$, effective anomalous Hall resistivity $\rho_{xy}^{A,\mathrm{eff}}$, and integrated low-energy spectral weight $W(E_F)$. Remarkably, all four quantities evolve on distinct temperature scales despite originating from the same correlated electronic system. In particular, the increase in low-energy $5f$ spectral weight sets in below approximately 90~K, near the broad maximum of $\rho_{xy}^{A,\mathrm{eff}}$, while neither quantity exhibits an abrupt change at $T_C$. These observations demonstrate that magnetic order and the reconstruction of the correlated electronic structure are intimately coupled but remain distinct phenomena. 

The different temperature dependencies further indicate that the anomalous Hall response cannot be viewed simply as a consequence of magnetic order acting on an otherwise fixed electronic structure. Instead, the anomalous Hall response continues to evolve while the correlated uranium $5f$ electronic structure itself is still undergoing substantial reconstruction. This conclusion is supported independently by ARPES, which directly reveals a continuous redistribution of low-energy spectral weight, and by the inability of magnetic skew scattering alone to account for the measured Hall response below the Curie temperature. Rather than being governed solely by the ordered magnetic moment, the anomalous Hall effect reflects the progressive evolution of the underlying correlated electronic structure. 

The first-principles calculations provide a microscopic connection between the topological electronic structure and the anomalous Hall response. Ferromagnetic order generates a symmetry-protected Weyl crossing approximately 0.13~eV above the Fermi level, accompanied by a pronounced sign-changing Berry-curvature distribution [Fig.~\ref{fig5}(e)]. Integration of the Berry curvature over the Brillouin zone yields $\sigma_{xy}^{\mathrm{int}}(E_F)\simeq9.6\times10^{2}~\Omega^{-1}\mathrm{cm}^{-1}$, of the same order of magnitude as the measured low-temperature value of approximately $4.5\times10^{2}~\Omega^{-1}\mathrm{cm}^{-1}$. This agreement in scale supports a substantial intrinsic contribution to the anomalous Hall response. The experimental results, however, demonstrate that the transport response continues to evolve after the magnetic ground state has already formed. This behavior suggests that the low-energy electronic states responsible for the anomalous Hall conductivity are progressively renormalized by electronic correlations within the ordered phase. The large anomalous Hall conductivity observed in UPS is therefore consistent with an interplay between nontrivial band topology and the temperature evolution of the correlated uranium $5f$ electronic states. Strong temperature-dependent evolution of coherent $5f$ states within magnetically ordered phases has also been observed in the antiferromagnetic uranium compounds USb$_2$ and UAs$_2$ \cite{Chen2019,Ji2022}. Recent studies of USbTe and UBiTe have likewise revealed large anomalous Hall conductivities associated with correlated uranium $5f$ electronic states \cite{Siddiquee2023,Xu2024}. In both systems, the intrinsic anomalous Hall response has been closely linked to the development of Kondo coherence. UPS exhibits a different phenomenology: no pronounced coherence maximum is observed in the electrical resistivity above the magnetic transition, while ARPES demonstrates that the low-energy electronic structure continues to evolve well inside the ferromagnetic state. UPS therefore represents a distinct regime in which magnetic order and correlated electronic reconstruction develop over overlapping temperature ranges rather than as separate sequential crossovers. Taken together, these observations place UPS in a regime where magnetic order, electronic correlations, and band topology evolve cooperatively, so that the anomalous Hall response reflects not only the Berry-curvature landscape but also the continuing reconstruction of the uranium $5f$ electronic structure.\\

\textit{Conclusions} - In summary, we have combined Hall transport, ARPES, thermodynamic measurements, and first-principles calculations to establish UPS as a correlated uranium Weyl semimetal. Resonant ARPES reveals the dual itinerant-correlated character of the uranium $5f$ electrons, while first-principles calculations identify a symmetry-protected Weyl crossing with pronounced Berry curvature and yield an intrinsic anomalous Hall conductivity of approximately $960~\Omega^{-1}\mathrm{cm}^{-1}$, comparable in scale to experiment. Most importantly, the anomalous Hall response does not simply track long-range ferromagnetic order: the low-energy $5f$ electronic structure continues to reconstruct well inside the ordered state. These results demonstrate a staged evolution of magnetism, correlated electronic structure, and topological transport in UPS, and establish uranium pnictochalcogenides as a promising platform for exploring how strong electronic correlations reshape topological phenomena.\\

\textit{Acknowledgments} - K.G., S.R., and S.Z. acknowledge support by the U.S. Department of Energy, Basic Energy Sciences, Materials Sciences and Engineering Division. C.K.S.\ and P.M.O.\ acknowledge support by the Knut and Alice Wallenberg Foundation (Grants No.\ 2022.0079 and No.\ 2023.0336). The calculations were partially supported by resources provided by the National Academic Infrastructure for Supercomputing in Sweden (NAISS) at NSC Link\"oping, partially funded by the Swedish Research Council through Grant No.\ 2022-06725. This research used resources of the Advanced Light Source, which is a DOE Office of Science User Facility under contract no. DE-AC02-05CH11231

\end{document}